\documentclass[twocolumn,tighten,times,twocolappendix,numberedappendix]{aastex631}

\usepackage{amsmath}
\usepackage{color}
\usepackage{url}
\usepackage{stmaryrd}

\usepackage{etoolbox}

\usepackage{setspace} 
\usepackage{graphicx}
\usepackage[flushleft]{threeparttable}
\usepackage{placeins}
\usepackage{comment}

\usepackage[encapsulated]{CJK}
\usepackage{ucs}
\usepackage[utf8x]{inputenc}

\hypersetup{breaklinks=true,colorlinks=true,linkcolor=blue,citecolor=blue,filecolor=blue,urlcolor=blue}
\makeatletter
\patchcmd{\NAT@citex}
  {\@citea\NAT@hyper@{%
     \NAT@nmfmt{\NAT@nm}%
     \hyper@natlinkbreak{\NAT@aysep\NAT@spacechar}{\@citeb\@extra@b@citeb}%
     \NAT@date}}
  {\@citea\NAT@nmfmt{\NAT@nm}%
   \NAT@aysep\NAT@spacechar\NAT@hyper@{\NAT@date}}{}{}

\patchcmd{\NAT@citex}
  {\@citea\NAT@hyper@{%
     \NAT@nmfmt{\NAT@nm}%
     \hyper@natlinkbreak{\NAT@spacechar\NAT@@open\if*#1*\else#1\NAT@spacechar\fi}%
       {\@citeb\@extra@b@citeb}%
     \NAT@date}}
  {\@citea\NAT@nmfmt{\NAT@nm}%
   \NAT@spacechar\NAT@@open\if*#1*\else#1\NAT@spacechar\fi\NAT@hyper@{\NAT@date}}
  {}{}
  
\usepackage{array}
\newcolumntype{C}[1]{>{\centering\let\newline\\\arraybackslash\hspace{0pt}}m{#1}}

\usepackage{xspace}

\def\aj{AJ}

\def\apj{ApJ}
\def\apjl{ApJ}
\def\apjs{ApJS}
\def\ao{Appl.~Opt.}

\def\aap{A\&A}
\def\aapr{A\&A~Rev.}
\def\aaps{A\&AS}

\def\mnras{MNRAS}

\def\pasp{PASP}

\def\procspie{Proc.~SPIE}
\def\rmxaa{Rev.~Mex.~Astron.~Astrofis.}

\def\arcdeg{\hbox{$^\circ$}}

\definecolor{burgundy}{rgb}{0.5, 0.0, 0.13}

\usepackage{xspace}

\newcommand{\orcidicon}{\includegraphics[width=0.26cm]{orcid-ID.eps}}

\newcommand{\orcidauthor}[1]{\href{https://orcid.org/#1}{\orcidicon}}

\shorttitle{Common envelope origins of the PN M\,3-38}
\shortauthors{Rechy-Garc\'{i}a et al.}

\makeatletter
\patchcmd{\frontmatter@RRAP@format}{(}{}{}{}
\patchcmd{\frontmatter@RRAP@format}{)}{}{}{}
\renewcommand\Dated@name{}
\makeatother

\begin{document}

\title{\large The common envelope origins of the fast jet in the planetary nebula M\,3-38}

\correspondingauthor{J.~S.\,Rechy-Garc\'{i}a}
\email{j.rechy@irya.unam.mx}

\author[0000-0002-0121-2537]{J.~S.\,Rechy-Garc\'{i}a}
\affil{Instituto de Radioastronom\'{i}a y Astrof\'{i}sica, UNAM, Antigua Carretera a P\'{a}tzcuaro 8701, Ex-Hda. San Jos\'{e} de la Huerta, Morelia 58089, Mich., Mexico}

\author[0000-0002-5406-0813]{J.~A.\,Toal\'{a}}
\altaffiliation{Visiting astronomer at the Instituto de Astrof\'{i}sica de Andaluc\'{i}a (IAA-CSIS, Spain) as part of the Centro de Excelencia Severo Ochoa Visiting-Incoming programme.}
\affil{Instituto de Radioastronom\'{i}a y Astrof\'{i}sica, UNAM, Antigua Carretera a P\'{a}tzcuaro 8701, Ex-Hda. San Jos\'{e} de la Huerta, Morelia 58089, Mich., Mexico}

\author[0000-0002-7759-106X]{M.~A.\,Guerrero}
\affil{Instituto de Astrof\'{i}sica de Andaluc\'{i}a, IAA-CSIC, 
Glorieta de la Astronom\'{i}a S/N, E-18008 Granada, Spain}

\author[0000-0001-5559-7850]{C.\,Rodr\'{i}guez-L\'{o}pez}
\affil{Instituto de Astrof\'{i}sica de Andaluc\'{i}a, IAA-CSIC, 
Glorieta de la Astronom\'{i}a S/N, E-18008 Granada, Spain}

\author[0000-0003-0242-0044]{L.\,Sabin}
\affil{Instituto de Astronom\'{i}a, UNAM, Apdo. Postal 877, Ensenada 22860, B.C., Mexico}

\author[0000-0003-2653-4417]{G.\,Ramos-Larios}
\affil{Instituto de Astronom\'{i}a y Meteorolog\'{i}a, CUCEI, Universidad de Guadalajara, Av. Vallarta 2602, Arcos Vallarta, 44130 Guadalajara, Mexico}

\date[]{Submitted to ApJL}

\begin{abstract}
We present the analysis of Multi-Espectr\'ografo en GTC de Alta Resoluci\'on para Astronom\'ia (MEGARA) high-dispersion integral field spectroscopic observations of the bipolar planetary nebula (PN) M\,3-38. 
These observations unveil the presence of a fast outflow aligned with the symmetry axis of M\,3-38 that expands with a velocity up to $\pm$225~km~s$^{-1}$. 
The deprojected space velocity of this feature can be estimated to be $\approx$320$^{+130}_{-60}$~km~s$^{-1}$, which together with its highly collimated morphology suggests that it is one of the fastest jet detected in a PN. 
We have also used Kepler observations of the central star of M\,3-38 to unveil variability associated with a dominant period of 17.7~days. 
We attribute this to the presence of a low-mass star with an orbital separation of $\approx$0.12--0.16~au. 
The fast and collimated ejection and the close binary system point towards a common envelope formation scenario for M\,3-38.
\end{abstract}


\keywords{\href{https://astrothesaurus.org/uat/1249}{Planetary nebulae (1249)};
\href{http://astrothesaurus.org/uat/1607}{Stellar jets (1607)};
\href{http://astrothesaurus.org/uat/1636}{Stellar winds (1636)};
\href{http://astrothesaurus.org/uat/2050}{Low mass stars(2050)}
\vspace{4pt}
\newline
}


\section{Introduction}
\label{wc:sec:introduction}

Planetary nebulae (PNe) are formed by mass lost by solar-like stars on their way of becoming white dwarfs, mostly during the asymptotic giant branch (AGB) phase \citep{Vassiliadis1993}. 
The resultant stellar remnant becomes hot enough to produce a high UV flux and fast stellar wind that ionizes and compresses the material ejected in the late AGB phase, giving birth to a PN \citep{Kwok2000}.

It has long been suggested that this generalized interacting stellar winds model of PN formation \citep[][]{Balick1987} is too simple to explain the wild variety of PN morphologies \citep[e.g.,][]{SCM1992,MGSS1996}. 
Several mechanisms were soon invoked to explain this conundrum \citep{Soker1997}, involving binary star interactions, including jet-like ejections during the late AGB or early post-AGB phases \citep{Sahai1998}.

Jets or bipolar fast outflows have been known to exist in PNe for a long time since the first discovery in NGC\,2392 \citep{Gieseking1985}. 
The statistical analysis of a sample of 85 collimated outflows detected in 58 PNe \citep[][and references therein]{Guerrero2020} found a bimodal distribution, with $\approx$30\% of collimated outflows showing velocities $\geq$100 km~s$^{-1}$. These high-velocity jets are prime candidates to have been launched by a binary system at the core of the PN.

The presence of fast collimated outflows can be gleaned by observing images of PNe as their action is typically associated with highly-elongated structures, S-shaped filaments and V-shaped structures \citep{Rechy2020a}.
Complementary kinematical information has been traditionally obtained using long-slit high-dispersion spectroscopic observations, which is otherwise the only method to detect fast collimated outflows in compact PNe \citep{Rechy2017}. 
The advent of high-dispersion integral field spectroscopy provides kinematical information with broad spatial coverage.  
This technique has allowed, for instance, to map for the first time the extension of the extremely faint bipolar jet in NGC\,2392 \citep{Guerrero2021}, to dissect the abundances and kinematics of the born-again PN HuBi\,1 \citep{Rechy2020b,MM2022} and the spatio-kinematical structure of M\,2-31 \citep{Rechy2021}.

\begin{figure*}
\begin{center}
\hspace*{-1.5cm}
\includegraphics[width=1.18\textwidth]{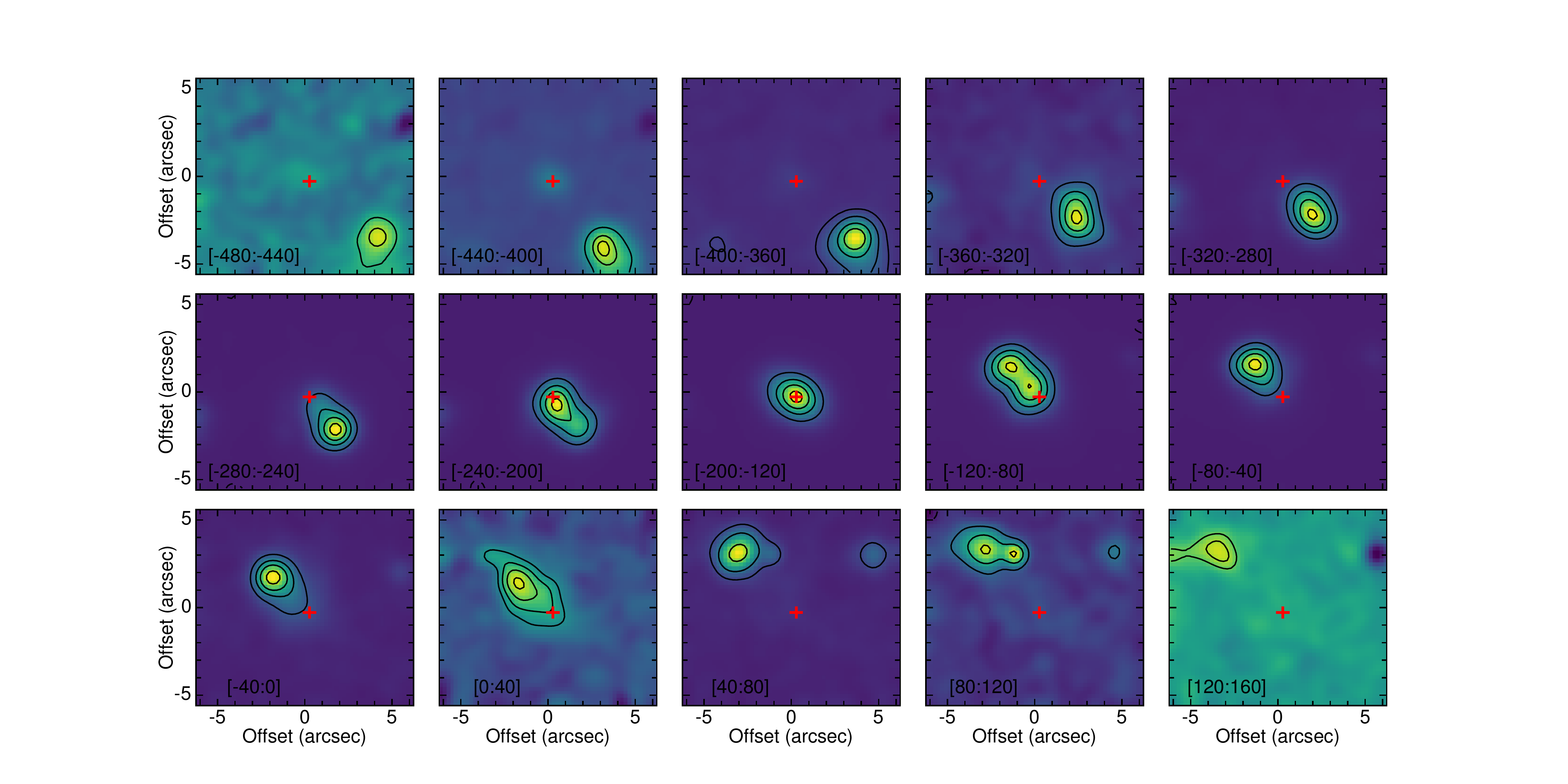}
\caption{GTC MEGARA tomography of the M\,3-38 in the [N~{\sc ii}] $\lambda$6584 emission line.  
The Local Standard Rest velocity range is labeled at the bottom-left corner of each panel.
The central frame corresponds to the systemic velocity.
The red ``cross'' indicates the location of the 
CSPN of M\,3-38, as determined from the emission of the stellar continuum, which is consistent with the central position of the nebular waist.  
North is up, east is to the left. 
}
\label{fig:M3_38_tomography}
\end{center}
\end{figure*}

In this letter, we present the discovery of one of the fastest collimated outflows in a PN. We use Multi-Espectr\'{o}grafo en GTC de Alta Resolución para Astronomía \citep[MEGARA;][]{GildePaz2018} observations obtained at the Gran Telescopio Canarias (GTC) of the PN M\,3-38 (PN\,G356.9$+$04.4) to unveil its extreme kinematic structure. Its central star (CSPN) is further found to exhibit a Kepler/K2 light-curve consistent with a post-common envelope (post-CE) binary system, which can be blamed as responsible for the collimation of the fast outflow in the recent past. 
The observations and archival data analysis of M\,3-38 were motivated by the presence of extremely fast ejections revealed in a long-slit echelle spectrum on the H$\alpha$+[N\,{\sc ii}] emission lines available at the San Pedro M\'artir Spectroscopic Catalogue of PNe \citep{Lopez2012}, and by the highly elongated bipolar morphology displayed in a narrow-band HST WFPC2 F656N image \citep[see figure 14 in][]{GuzmanRamirez2014}.

This letter is organized as follows.  
In Section~\ref{sec:observations} we describe the observations and the data preparation, in Section~\ref{sec:results} we present our results, and finally the discussion is addressed in Section~\ref{sec:discussion}.

\section{Observations and data preparation}
\label{sec:observations}

M\,3-38 was observed with GTC MEGARA on 2021 August 29 using the Integral Field Unit (IFU) mode. 
This provides a field of view of 12\farcs5$\times$11\farcs3 with a spaxel size of 0\farcs62. 
We used the high-resolution (HR) Volume-Phased Holographic (VPH) grism VPH665-HR covering the 6405.6-6797.1~\AA\, wavelength range at a spectral dispersion of 0.098~\AA\ with a resolution of $R=$18,700 ($\simeq16$ km~s$^{-1}$). 
Three 900 s exposures were obtained with a seeing of 1\farcs2.

The data reduction process was carried out using the Data Reduction Cookbook provided by the Universidad Complutense de Madrid \citep[][]{Pascual2021}. Through the pipeline, we subtracted the sky and bias contributions, correct for flat field effects, and perform wavelength calibration, spectra tracing and extraction. Sky subtraction is performed via 56 ancillary fibers located $\approx2\farcm0$ from the center of the IFU. We apply the regularization grid task {\it megararss2cube}\footnote{Task developed by J.\,Zaragoza-Cardiel available at \url{https://github.com/javierzaragoza/megararss2cube}} to obtain square spaxels with sizes of 0\farcs215. The final cube has dimensiones of 52$\times$58$\times$4300 equivalent to a total of 3016 spectra in the datacube. The flux calibration was done using observations of the spectrophotometric standard star HR7596 taken immediately after the observations of M\,3-38.

\begin{figure*}
\begin{center}

\includegraphics[width=0.98\linewidth]{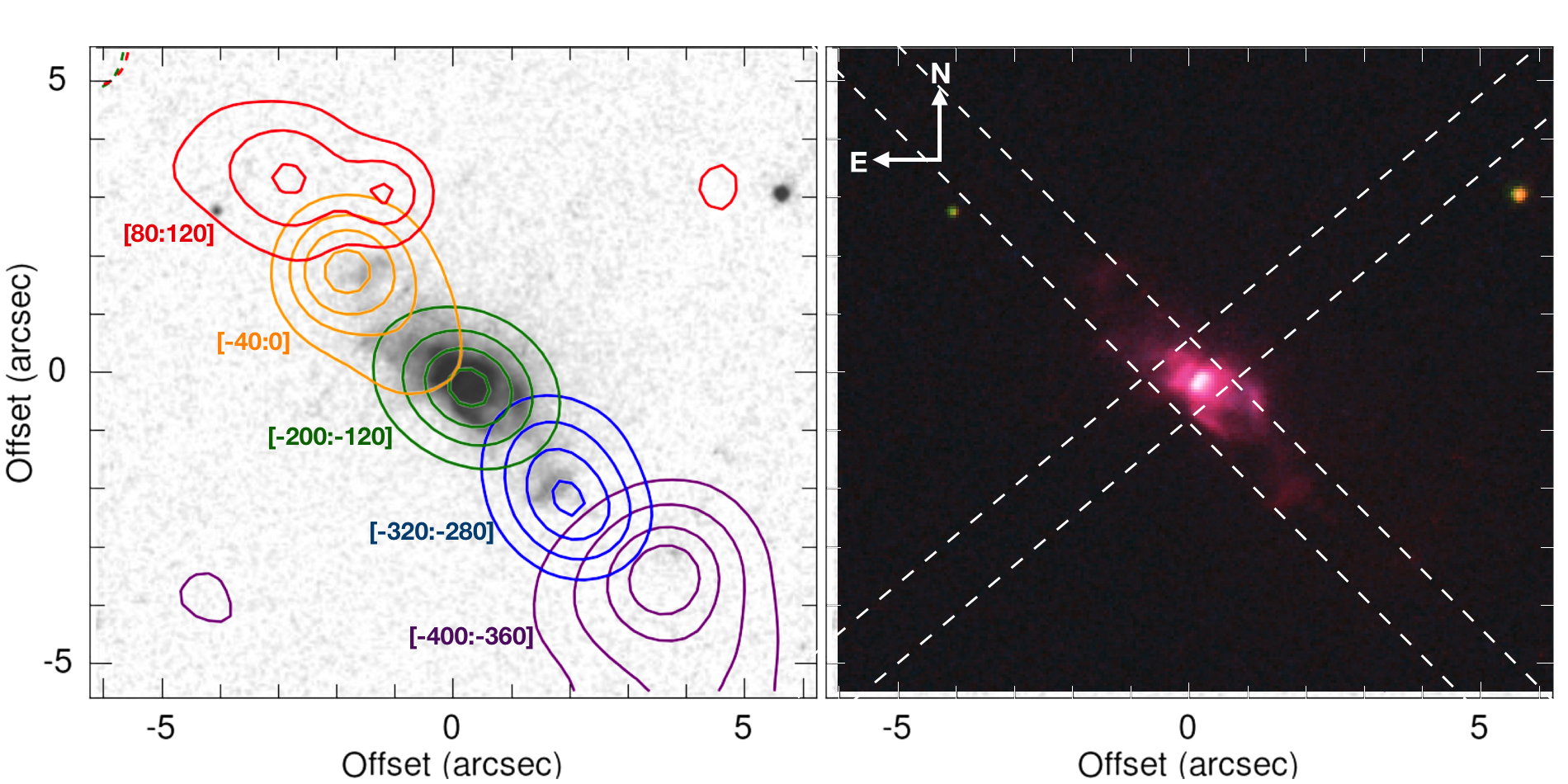}

\caption{
HST WFPC2 optical images of M\,3-38. 
The left panel shows a grey-scale F656N (H$\alpha$) image overimposed with contours from the MEGARA [N~{\sc ii}] $\lambda$6584~\AA\ emission line. 
The colors represent different velocity ranges as labeled. 
The right panel shows a color-composite image obtained by combining F656N H$\alpha$ (red), F547M (green), and F502N [O~{\sc iii}] (blue) images. 
The dashed-line areas represent the extraction regions for the PV diagrams shown in Figure~\ref{fig:PV}.
}
\label{fig:HST}
\end{center}
\end{figure*}

M 3-38 was observed by the Kepler/K2 mission \citep{Howell2014} during Campaign 11 in 2016 during September 24 to December 8. 
Campaign 11 suffered a three day interruption (October 18--20) to make a pointing correction of $-0.32^\circ$ in the spacecraft attitude. 
As a result, the dataset was splitted in two parts, designated Campaign 111 and 112, spanning 23 and 48 days, respectively (see Figure~\ref{fig:kepler_lc}\textbf{,} upper left), that were processed separately using new target apertures in the second part of the campaign (see the Data Release Notes\footnote{\url{https://archive.stsci.edu/missions/k2/doc/drn/KSCI-19151-002\_K2-DRN30\_C11.pdf}} for more information). 
We retrieved the two lightcurves from the Mikulski Archives for Space Telescopes\footnote{\url{https://mast.stsci.edu/portal/Mashup/Clients/Mast/Portal.html}} and used the \textit{Pre-search Data Conditioning Simple Aperture Photometry} (PDCSAP) datasets, already cleaned for systematic errors. 
After removing outliers, the flux was normalized and the two datasets joined together resulting in a total of 3166 useful points with a 30~minute cadence along a 74 days time baseline.

\section{Results}
\label{sec:results}

\subsection{Kinematics of the collimated outflow}

The MEGARA IFS observations of M\,3-38 indicate diffuse emission in the H$\alpha$, [N~{\sc ii}] $\lambda\lambda$6548,6584, [S~{\sc ii}] $\lambda\lambda$6717,6731, and He\,{\sc ii} $\lambda$6560 lines. 
Emission from the [N~{\sc ii}] $\lambda\lambda$6435.61,6527.23, C~{\sc i} $\lambda\lambda$6674.11,6683.95, and C~{\sc ii} $\lambda$6578.05 lines is also detected in the innermost regions.  
The spatio-kinematics of M\,3-38 will be investigated here using the brightest H$\alpha$ and [N~{\sc ii}] $\lambda$6584 emission lines.

\begin{figure*}
\begin{center}
\includegraphics[width=\linewidth, trim=85 0 50 0cm]{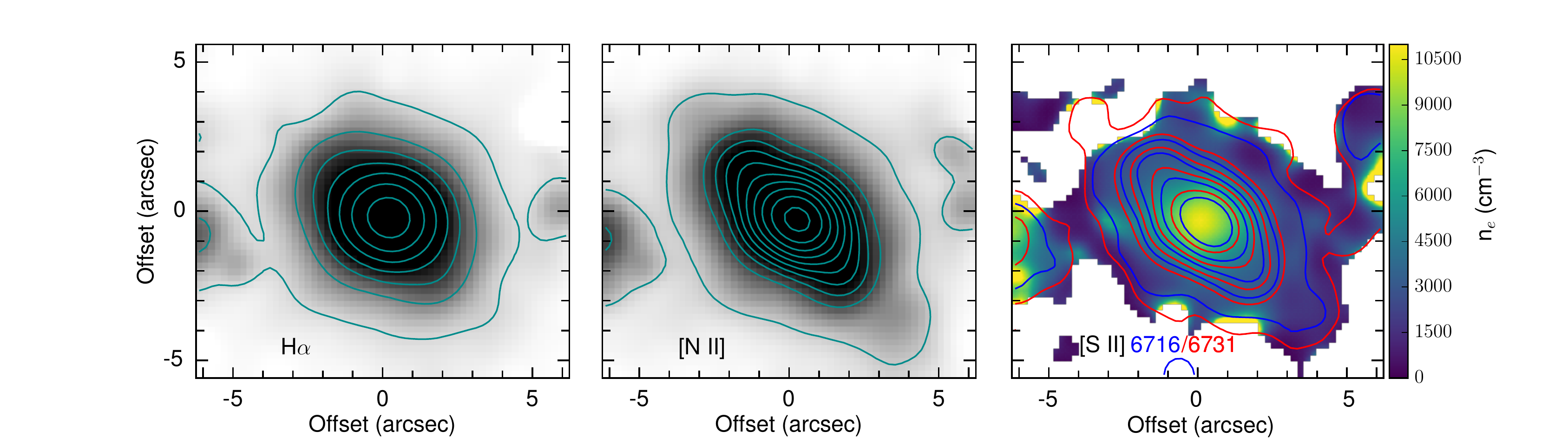}
\caption{
GTC MEGARA continuum-subtracted H$\alpha$ (left) and [N~{\sc ii}] $\lambda$6584 (centre) line emission maps, and electron density map derived from the [S~{\sc ii}] $\lambda\lambda$6716,6731 line ratio (right, see text for details).  
Contours are used to emphasize regions both of faint and bright surface brightness. 
The blue and red contours in the density map correspond to the same surface brightness levels in the [S~{\sc ii}] $\lambda$6716 and [S\,{\sc ii}] $\lambda$6731 images, respectively. 
}
\label{fig:Image}
\end{center}
\end{figure*}

\begin{figure*}
\begin{center}

\includegraphics[width=1\linewidth,trim=0 60 0 0cm]{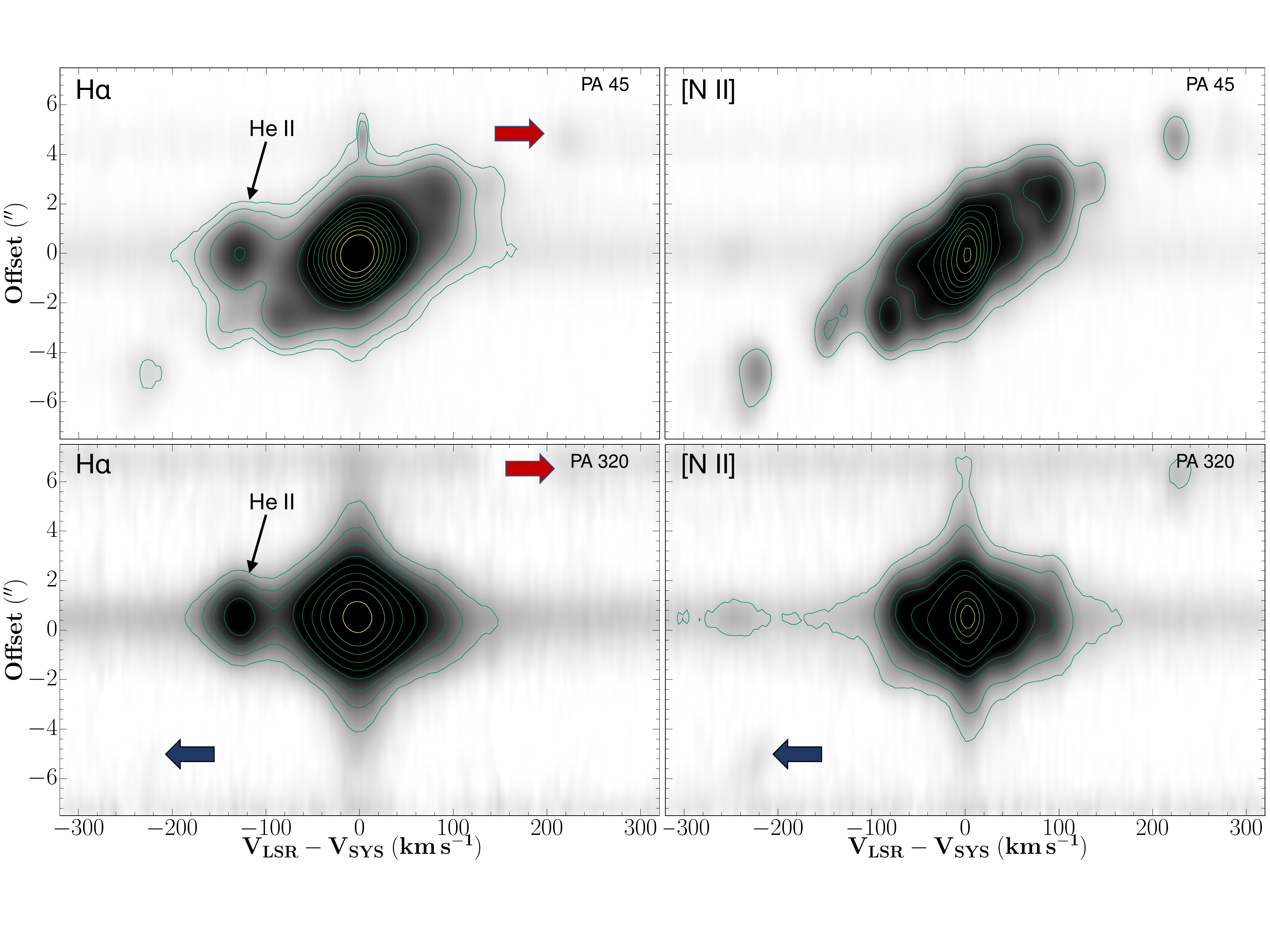}
\caption{PV diagrams of the H$\alpha$ (left) and [N~{\sc ii}] $\lambda$6584 (right) emission lines extracted from MEGARA pseudo-slits along the bipolar axis at PA 45$^{\circ}$ (top) and an orthogonal direction at PA 320$^{\circ}$ (bottom). 
Colored contours are used to emphasize both faint and bright emission. 
When the emission from high-velocity components (see text) is too faint to be shown by these contours, red and blues arrows are used to indicate receding and approaching features.
}
\label{fig:PV}
\end{center}
\end{figure*}

A tomographic view of M\,3-38 is displayed in Figure~\ref{fig:M3_38_tomography}, with channel maps of the [N~{\sc ii}] $\lambda$6584 emission line spanning from radial velocities in the range $-$480 to $+$160~km~s$^{-1}$ with a width of 40~km~s$^{-1}$. 
For a systemic velocity in the Local Standard Rest (LSR) frame of $v_\mathrm{sys}=-$160.5~km~s$^{-1}$ (central panel of Fig.~\ref{fig:M3_38_tomography}), the emission in the most extreme panels implies expansion velocities in the range $\pm$225 km~s$^{-1}$. 
The emission of the fast outflow of M\,3-38 is aligned along a position angle (PA) of $43^\circ\pm2^\circ$, with the southwest emission approaching (blue-shifted) and the northeast emission receding (red-shifted). 
The emission is thus mostly aligned with the symmetry axis of M\,3-38, as illustrated in the left panel of Figure~\ref{fig:HST}, where contours of the emission in different velocity ranges are overplotted on an HST WFPC2 H$\alpha$ image. 
In this image, M\,3-38 exhibits a highly axisymmetric morphology with two 5.3$''$ in size bipolar lobes aligned along a PA of $39^\circ\pm6^\circ$ terminating in bow-shock-like clumps. 
The emission from the intermediate-velocity ranges from 
$-40$ to 0 km~s$^{-1}$ (orange in Fig.~\ref{fig:HST}-{\it left}) and 
$-320$ to $-$280 km~s$^{-1}$ (blue in Fig.~\ref{fig:HST}-{\it left}) is associated with those clumps, but the emission from the extreme-velocity ranges from 
$+80$ to $+120$ km~s$^{-1}$ (red in Fig.~\ref{fig:HST}-{\it left}) and 
$-400$ to$-360$ km~s$^{-1}$ (violet in Fig.~\ref{fig:HST}-{\it left}) peaks at $\simeq5''$ from the CSPN, i.e., they extend beyond the emission detected in the HST H$\alpha$ image.

The extent of the emission detected by MEGARA is illustrated in the continuum-subtracted H$\alpha$ and [N~{\sc ii}] $\lambda$6584 maps presented in Figure~\ref{fig:Image}. 
A density map is also presented in the right panel of this figure.  
The density has been computed from the [S~{\sc ii}] $\lambda$6716 to [S~{\sc ii}] $\lambda$6731 ratio map using in python getTemDen function of the PyNeb package \citep{Luridiana2015} for the analysis of emission lines assuming an electron temperature\footnote{The calculation of the density using the [S~{\sc ii}] doublet is rather insensitive to the adopted value of electron temperature for typical values in the range 8000--12000 K found in PNe.} of 10000~K.
This map indicates a clear density gradient from the innermost regions of M\,3-38 to the inner outflow ($N_{\rm e} \approx 2,500$ cm$^{-3}$) and to its outermost tip ($N_{\rm e} \approx 1,500$ cm$^{-3}$).

To investigate the expansion-law of the outflow of M\,3-38, classical position-velocity (PV) diagrams in the H$\alpha$ and [N~{\sc ii}] $\lambda$6584 emission lines have been extracted from a pseudo long-slit along its bipolar axis at PA 45$^\circ$ (Fig.~\ref{fig:PV}-{\it top}). 
These PV diagrams are consistent with a homologous expansion with a velocity gradient $\approx$60 km~s$^{-1}$~arcsec$^{-1}$. 
We note that these PV diagrams are suggestive of a double-S shape that could be attributed to precession. 
The highest-velocity components at the tip of this outflow reach expansion velocities $\pm$220 km~s$^{-1}$, with the SW clump showing an intriguing hook-like shape extending up to $\approx$7$^{\prime\prime}$ from the CSPN with expansion velocity of $-$240 km~s$^{-1}$.

The MEGARA observations also detect another pair of high-velocity clumps, as denoted by the violet and red contours in Fig.~\ref{fig:HST}-{\it left}, located $\simeq5.5''$ from the CSPN towards the SE and NW of M3-38, respectively.  
Quite surprisingly, these clumps are aligned along a direction which is almost orthogonal to the bipolar axis of M\,3-38.
To illustrate further their presence, H$\alpha$ and [N~{\sc ii}] $\lambda$6584 PV diagrams have also been extracted from a $1^{\prime\prime}$-wide pseudo long-slit along PA=320$^\circ$ (Fig.~\ref{fig:HST}-{\it right}). 
These PV diagrams (Fig.~\ref{fig:PV}-{\it bottom}), and particularly the [N~{\sc ii}] one, confirm the presence of high-velocity components along this direction with an expansion velocity of $\pm$230 km~s$^{-1}$. 

Incidentally, we shall mention that the PV diagrams shown in Figure~\ref{fig:PV} reveal the presence of an extended component at the systemic velocity of M\,3-38 that seems to fill the whole field of view of the MEGARA IFU.  
This component is broad and unresolved, with an intrinsic FWHM of 40 km~s$^{-1}$ and a line-tilt up to $\simeq$6 km~s$^{-1}$.
It can thus be interpreted as a low-velocity non-spherical halo around M\,3-38.

\begin{figure*}
\begin{center}
\includegraphics[width=\linewidth]{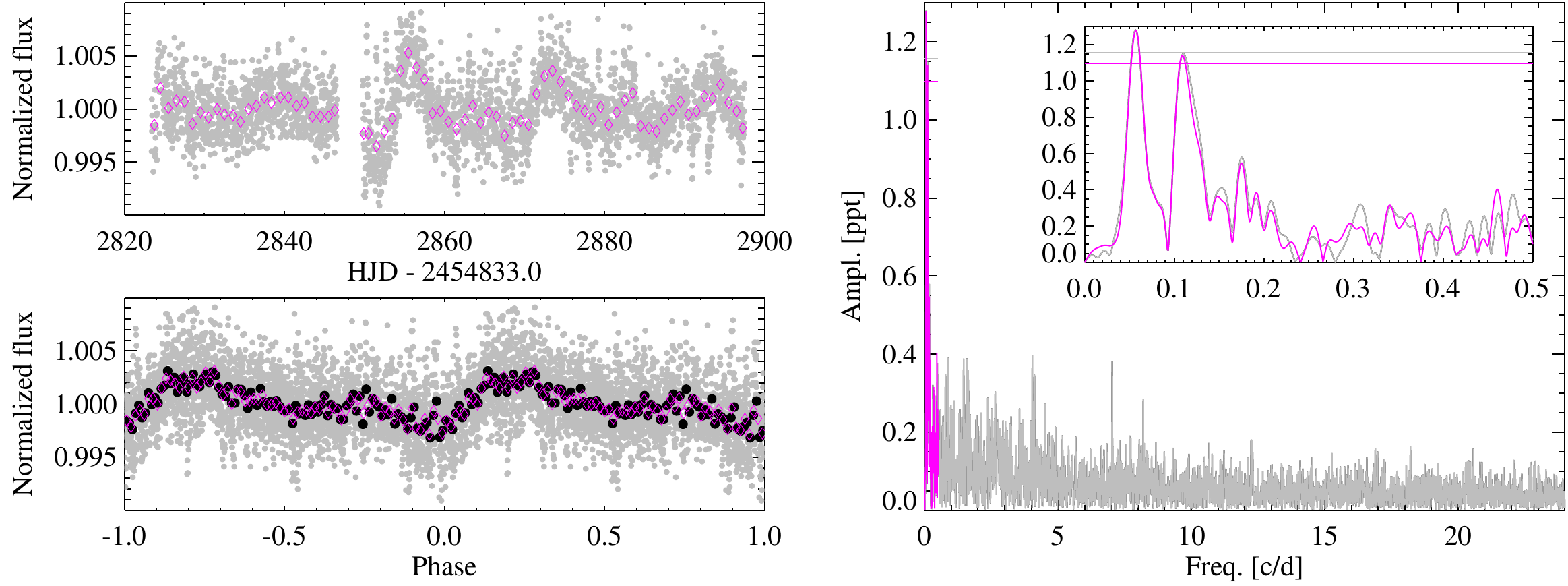}
\caption{
Top-left panel: original (grey) and one day (magenta) binned Kepler/K2 C111 and C112 light curve of the CSPN of M\,3-38. 
The two campaigns are separated by a three day gap. Bottom-left panel: light curve of the CSPN of M\,3-38 folded to 17.7 days (symbols are the same as in the upper plot). 
Black points correspond to 0.01-phase binned data. 
Right panel: amplitude spectra for the individual and one-day binned datasets. The horizontal lines in the amplitude spectra are four times the mean amplitude in a 0.5~c/d range around the main frequency. 
The insert zooms in the amplitude spectra up to the Nyquist frequency of the binned data series.}
\label{fig:kepler_lc}
\end{center}
\end{figure*}

\subsection{Photometric variations of the central star}

We used {\it Period04} \citep{Lenz2005} to search for periodicities in the Kepler/K2 light curve of the CSPN of M\,3-38 derived from Campaign C111 and C112 (Fig.~\ref{fig:kepler_lc}-{\it left-top}).
The amplitude spectra was calculated up to the Nyquist frequency, corresponding to 24~cycle~day$^{-1}$. 
A frequency is considered significant when its amplitude is at least four times the noise level \citep{Breger1993}, which is equivalent to a 99.9\% confidence level \citep{Kuschnig1997}. 
Since no significant frequencies were found in the high frequency range, the original light curve was binned to one day bins to search for periodicities up to the 0.5~cycle~day$^{-1}$ Nyquist limit (Fig.~\ref{fig:kepler_lc}-{\it top-left panel}).  
In both amplitude spectra (Fig.~\ref{fig:kepler_lc}-{\it right panel}), we found the highest amplitude frequency at 0.0565~cycle~day$^{-1}$ to be significant, corresponding to a main period of $\sim$17.7~d, and its first harmonic at 0.1199~cycle~day$^{-1}$, corresponding to $\sim$9.0~d period. 
We note that the analysis of the same Kepler/K2 data by \citet{Jacoby2021} deemed it non-variable, as these authors conservatively excluded cases of low-amplitude variability if the best period was close to or more than half of the sampling window that could have resulted from spurious variability. 

Fig.~\ref{fig:kepler_lc}-{\it bottom-left panel} shows the folded individual and one day binned light curve to the main frequency, as well as mean values for a 0.01 phase bins for better appreciation. 
The light curve shows a dominant peak at phase 0.2 with a main minimum at phase 0.0 and a subtle through at phase 0.55. Within the limits of the low signal, there seem to be a hint of a secondary peak at phase 0.75.
The CSPN of M\,3-38 is classified as an H-deficient, weak emission line star \citep{DePew2011}, but the coherence of its light-curve precludes wind variability \citep[see, for instance, the light-curve of H\,2-48 in Figure~3c of][]{Jacoby2021}, whereas long-lived bright spots on the stellar surface would have resulted in shorter periods associated with the stellar rotation.
The variations in the light-curve of M\,3-38 could be attributed to irradiation onto a companion, as an ellipsoidal modulation due to tidal distortion would require a much shorter period.

\section{Discussion}
\label{sec:discussion}

In the previous section we have discussed the presence of a fast collimated outflow and a binary star in M\,3-38.  
Their properties are discussed below.

The main outflow of M\,3-38 is aligned along its bipolar axis and has expansion velocities up to $\approx$225~km~s$^{-1}$ (Fig.~\ref{fig:HST} and \ref{fig:PV}). 
This makes the outflow in M\,3-38 one of the fastest among PNe, only surpassed by those in M\,1-16 and MyCn\,18 and as fast as that in KjPn\,8 \citep[][and references therein]{Guerrero2020}. 
The high velocity, linear PV diagram, and elongated morphology of this outflow makes it a ``bona-fide'' jet candidate. 
The jet space velocity cannot be derived because its inclination angle with the line of sight is unknown.  
If an inclination of 45$^\circ\pm15^\circ$ were to be adopted, then the space velocity of the jet would be $320^{+130}_{-60}$~km~s$^{-1}$. The outermost fastest component of this jet is projected about 5$''$ from the central star of M\,3-38 (Fig.~\ref{fig:HST}). 
Adopting the most recent distance estimate of 8~kpc to M\,3-38\footnote{
Distance estimates to M\,3-38 in the literature range from 5.3 to 14.1 kpc, with an average value very similar to the one adopted in the text.
} 
derived from MSX MIR flux densities by \citet{Ortiz2011} and a similar $45^\circ\pm15^\circ$ inclination angle, the projected distance from the PN center of 0.19 pc would imply a true linear radius of 0.27$^{+0.11}_{-0.05}$ pc. This translates into a jet kinematic age of  860$^{+630}_{-360}$~yr.  
We remark that the values of the projected distance, true linear radius, and kinematic age given above scale with the distance, which has been adopted to be 8 kpc.

The ionized mass of the jet can be computed from the relationship \citep{pottasch1984}:
\begin{equation}
M_\mathrm{ion}\,[\mathrm{M}_\odot] = 11.06 \times F(\mathrm{H}\beta)\frac{d^2(T_\mathrm{e}/10^4)^{0.88}}{N_\mathrm{e}},
\end{equation}
where the intrinsic H$\beta$ flux is in units of $10^{-11}$ erg~cm$^{-2}$~s$^{-1}$ and $d$ in kpc.  
The intrinsic H$\beta$ flux of the jet is derived to be $4.2\times10^{-12}$ erg~cm$^{-2}$~s$^{-1}$ from the observed H$\alpha$ flux of $5.3\times10^{-13}$ erg~cm$^{-2}$~s$^{-1}$ for a logarithmic extinction $c$(H$\beta$) of 2.0 \citep{Tylenda1992} and case B recombination.  
The density of the jet is $\simeq$2,500 cm$^{-3}$, as derived from the density-sensitive [S~{\sc ii}] doublet (Fig.~\ref{fig:Image}-right). 
Assuming a value of 10,000 K for $T_{\rm e}$, an ionized mass of 0.011~M$_\odot$ is derived. 
Alternatively, the volume occupied by the jet and its density can be used to estimate the ionized mass assuming a conservative value of 0.1 for the ``filling factor'' $\epsilon$ and a standard value of 1.4 for the mean molecular weight. 
Adopting a single cylindrical geometry with radius 0.5$''$ and height 5$''$ tilted by $45^\circ\pm15^\circ$ with the line of sight, a value of $\approx3\times$10$^{-3}$ M$_\odot$ is derived.
Given the uncertainties and assumptions of both methods, we shall adopt an intermediate value for the ionized mass of $M_\mathrm{ion}=6\times$10$^{-3}$ M$_\odot$.

The values of space velocity\footnote{
An average radial expansion velocity of 70~km~s$^{-1}$, which is representative of the bulk of emission, was adopted in the calculations of the linear momentum and mechanical luminosity, rather than the extreme 225~km~s$^{-1}$ expansion velocity.}, kinematic age, and ionized mass can be combined to derive a mass-loss rate in the range $2\times10^{-6}-2\times10^{-5}$~M$_\odot$~yr$^{-1}$, 
a linear momentum from $6\times10^{34}$ to $2.8\times10^{35}$~dyn, 
and a mechanical luminosity of 
$2-20$~L$_\odot$ for the bulk of the emission.

The analysis of the Kepler/K2 light-curve reveals a dominant frequency that corresponds to a period of 17.7~d, suggesting that M\,3-38 host a binary system. 
The variability type is unclear as it can be assigned either to irradiation of a main sequence companion \citep{Hillwig2016a} or to ellipsoidal modulation, which is always interpreted in CSPNe as a case for a double degenerate system \citep[DD;][]{Hillwig2010}. 
By adopting a total mass for the binary CSPN of M\,3-38 of 0.7--1.6 M$_{\odot}$ (0.4-1.0 M$_{\odot}$ for the CSPN and 0.3-0.6 M$_{\odot}$ for the companion whatever it is a WD or a main sequence star), an orbital separation of $\approx$0.12--0.16~au ($\approx$25--34~R$_{\odot}$) is estimated.  
This orbital separation favors irradiation as a viable mechanism to produce the small amplitude variation seen in M\,3-38 \citep[see the top-left panel of figure~1 in][]{DMetal2008}, whereas it makes ellipsoidal variability highly unlikely for the expected Roche lobe size of several solar radii.

The small orbital separation and the presence of highly-collimated, fast ejections from M\,3-38 leads us to suggest that it might have formed as the result of CE evolution \citep[e.g.,][]{Paczynski1976,Ivanova2013,Jones2017}. 
During the CE phase the massive component overflows its Roche lobe and the companion is engulfed. 
Models predict that the orbit decays and orbital energy and angular momentum are transferred to the CE within a timescale of months-to-years \citep[e.g.,][]{Chamandy2020}. 
The subsequent evolution naturally explains the formation of bipolar ejections \citep[see, e.g.,][and references therein]{GS2021,Ondratschek2021,Zou2020,LopezCamara2021} of material in the direction perpendicular to the orbital plane \citep[see][]{Hillwig2016}.  
Indeed, high-resolution ALMA observations of water fountain objects, which are often interpreted as transitional sources between the AGB and proto-PNe phases, imply that they have experienced a recent ($<$200~yr) CE evolution \citep{Khouri2021}.

Mass estimates and kinematics of PNe that experienced a CE phase can be directly contrasted with theoretical predictions \citep{Alcolea2007,Frew2007,Corradi2015,SantanderGarcia2022}. 
The mass-loss rate of the jets of M\,3-38 is about one to two orders of magnitude larger than those reported for the jets of PNe with post-CE close binaries \citep{Tocknell2014} and for the jet in NGC\,2392 \citep{Guerrero2021}, but consistent with those of the late AGB jets in BD$+$46$^\circ$442 and IRAS\,19135$+$3937 reported by \citet{Bollen2020}.  
The linear momentum and mechanical luminosity of the jet in M\,3-38 are in the high end of the range presented by those authors, but still notably smaller than the values reported for outflows in proto-PNe \citep{Bujarrabal2001}.

If the jet in M\,3-38 were launched by a CE interaction, its velocity should be of the order of the escape velocity of the secondary star launching the jet,
\begin{equation}
    v_\mathrm{j} = \sqrt{\frac{2 G M}{R}},
\end{equation}
\noindent 
where $M$ and $R$ are its mass and radius, and $G$ is the gravitational constant. 
Jet velocities of 540 and 480~km~s$^{-1}$ can be estimated for M5\,V and M8\,V stars with masses of 0.21 and 0.06~M$_\odot$, respectively \citep[see][]{Cox2000}. These values are close to the upper limit estimated for the real jet velocity, but we note that the interaction of the jet with the material ejected previously during the AGB phase might have slowed it down. 

The 17.7~d period proposed for the binary CSPN of M\,3-38 is relatively long compared to the majority of binary CSPNe \citep[see the compilation presented by][and references therein]{BJ2019}. In contrast, simulations of CE evolution result in relatively large separations of the order of 10--20~R$_\odot$ and periods larger than 3~d \citep[see, e.g.,][]{Chamandy2018,Chamandy2020,Iaconi2019}, more alike that of M\,3-38.

To conclude, M\,3-38 could be a new addition to the select group of PNe with fast collimated outflows and post-CE CSPNe \citep[e.g., Fleming\,1,][]{Boffin2012}. 
A more accurate characterization of the light-curve of its CSPN and properties of its jet are urgent, as M\,3-38 could be a case study of the late evolution of low- and intermediate-mass stars in binary systems and the launch of high-velocity jets in a CE phase. 
Its unusually long orbital period among binary post-CE CSPNe makes it also a key case for the comparison of CE evolution simulations and observations. 

Finally, it is interesting to remark the presence of an additional high-velocity ejection almost orthogonal to the main one.  
While astonishing, this is not uncommon among PNe, with examples of collimated outflows almost orthogonal \citep[e.g.\ IC\,4593;][]{Corradi1997} or along very different directions \citep[see the compilation presented by][]{Guerrero2021}, with the record case of NGC\,6210 with five different symmetry axes \citep{Henney2021}. 
It has been noted that jets misaligned with the main nebular axis might be characteristic of PNe with a post-CE binary \citep{BondLivio1990,DeMarco2009}.  
There is not an obvious interpretation for these phenomena \citep[see, e.g.,][]{BearSoker2017}, but, if associated with a CE phase, it is clearly suggestive of dramatic changes in the preferential ejection direction of the stellar system. 
\\


\noindent JSRG acknowledges support from the Programa de Becas posdoctorales of the Direccion General de Asuntos del Personal Acad\'{e}mico of the Universidad Nacional Autonoma de M\'{e}xico (DGAPA, UNAM, Mexico). JAT acknowledges support by the Marcos Moshinsky Foundation (Mexico), UNAM PAPIIT project IA101622 (Mexico), and the Visiting-Incoming programme of the Centro de Excelencia Severo Ochoa (Spain). MAG acknowledges support of grant PGC 2018-102184-B-I00 of the Ministerio de Educaci\'on, Innovaci\'on y Universidades cofunded with FEDER funds. CR-L and MAG acknowledge financial support from the State Agency for Research of the Spanish MCIU through the ``Center of Excellence Severo Ochoa" award to the Instituto de Astrof\'\i sica de Andaluc\'\i a (SEV-2017-0709). 
LS acknowledges support from UNAM PAPIIT Grant IN110122. JAT AND GR-L acknowledges support from CONACyT (grant 263373). The authors are thankful to the referee, Prof. Orsola De Marco, for comments and suggestions that improved the interpretation of the results presented here, to E.\ Rodr\'\i guez for helpful discussions on the binary nature of the CSPN of M\,3-38, and to G.\ Jacoby and D.\ Jones for discussions and further clarification of the methods and results presented in their work \citep{Jacoby2021}. This work has made extensive use of the NASA’s Astrophysics Data System.



\software{megararss2cube}
\facilities{Gran Telescopio Canarias (MEGARA), Kepler/K2, HST (WFPC2)}




\end{document}